# A New Health Strategy to Prevent Pressure Ulcer Formation in Paraplegics using Computer and Sensory Substitution via the Tongue

Alexandre MOREAU-GAUDRY[a,c,1], Anne PRINCE[b],
Jacques DEMONGEOT[a,c] and Yohan PAYAN[c]

[a] *University Hospital of Grenoble, France*
[b] *University and Medical Center of St Hilaire du Touvet, France*
[c] *TIMC-IMAG-CNRS Laboratory, Grenoble, France*

**Abstract.** Pressure ulcers are recognized as a major health issue in individuals with spinal cord injuries and new approaches to prevent this pathology are necessary. An innovative health strategy is being developed through the use of computer and sensory substitution via the tongue in order to compensate for the sensory loss in the buttock area for individuals with paraplegia. This sensory compensation will enable individuals with spinal cord injuries to be aware of a localized excess of pressure at the skin/seat interface and, consequently, will enable them to prevent the formation of pressure ulcers by relieving the cutaneous area of suffering. This work reports an initial evaluation of this approach and the feasibility of creating an adapted behavior, with a change in pressure as a response to electro-stimulated information on the tongue. Obtained during a clinical study in 10 healthy seated subjects, the first results are encouraging, with 92% success in 100 performed tests. These results, which have to be completed and validated in the paraplegic population, may lead to a new approach to education in health to prevent the formation of pressure ulcers within this population.

## 1. Introduction

A pressure ulcer is defined as an area of localized damage to the skin and underlying tissue caused by pressure, shearing, friction or a combination of these factors [1]. Its prevalence, which ranges from 23% to 39% in adults with spinal cord injuries (as reported in two recent retrospective and one prospective studies), remains high in this population [2-4]. Located near bony prominences such as the ischium, sacrum and trochanter, pressure ulcers are recognized as the main cause of rehospitalization for patients with paraplegia [5]. Their treatment, which can be medical or surgical, is always long, difficult and expensive. Thus, this pathology appears to be a major health issue for this population.

[1] Corresponding Author: Alexandre Moreau-Gaudry. TIMC-IMAG-CNRS Laboratory. GMCAO Team. Faculty of medicine, Alexandre.Moreau-Gaudry@imag.fr

Because of the need to identify effective intervention strategies that provide health education for skin care management and the prevention of pressure ulcers for individuals with paraplegia, a new approach has been developed using computer-aided instruction as an educational tool for promoting independent skin care [6]. As with traditional approaches, the main purpose of this strategy is to control the pressure applied at the seat/skin interface in order to prevent the formation of pressure ulcers. Nevertheless, to our knowledge, no approach has been developed that compensates for the sensory loss for paraplegics in the buttock area which would allow them to lead a similar lifestyle to that of healthy subjects in term of pressure ulcer prevention. Healthy subjects prevent this pathology by moving the body in a conscious or subconscious manner according to a "signal" arising from the buttock area. The main purpose of this research is to define a new health strategy that prevents the formation of pressure ulcers in individuals with paraplegia by simulating the natural performance of a healthy subject via use of sensory substitution. The first results related to the feasibility of this approach in healthy seated subjects are reported in this work.

## 2. Material and Methods

### 2.1. Which organ is best suited to sensory substitution?

The concept of sensory substitution has its origins in the works of P. Bach-y-Rita, who has studied this area for 30 years and, in particular, sensory substitution for blind people. Indeed, he states: "we do not SEE with the eyes". The visual image does not go beyond the retina, but is turned into patterns of pulses along nerves and is carried to the brain [7]. To demonstrate this theory, a human-machine interface was developed: the Tactile Vision Substitution System (TVSS), which allows visual information to be emitted from a TV camera to an array of stimulators in contact with the skin on one or several parts of the body (abdomen, back, and forehead). After training with this device, blind people are able to perform complex perceptual and "eye"-hand coordination tasks. Furthermore, when the human-machine interface is moved from one area of skin to another, there is no loss of correct spatial localization. Thus, the trained blind subject does not perceive the image on the skin, but in fact locates it correctly in space.

More recently, P. Bach-y-Rita has focused on the tongue as the most ideally suited organ for the human-machine interface because of outstanding tactile and physical properties. He also reported the possibility of developing a cosmetically acceptable interface into an orthodontic retainer thanks to recent technological improvements [8]. For these reasons, and because the sensory and motor nervous pathways for the tongue are preserved in paraplegia, this organ was chosen for sensory substitution in spinal cord injury individuals.

### 2.2. The implementation of sensory substitution.

A new device has been developed to compensate for sensory loss in paraplegics in the buttock area. It consists of three components (Figure 1). The first is a pressure mapping system which allows real-time acquisition of the pressure applied on the seat/skin interface. The second is the human-machine interface. In a preliminary experiment, this device was the Tongue Display Unit (TDU) developed by P. Bach-y-Rita and colleagues.

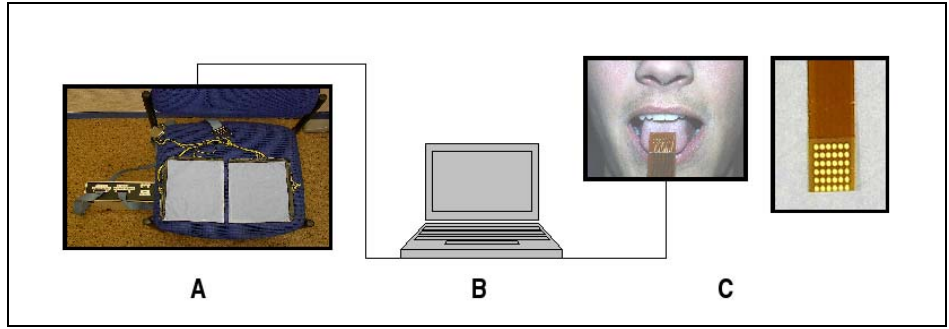

**Figure 1.** A new device has been developed, attempting to compensate for sensory loss in paraplegics. The pressure mapping system (A) allows real-time acquisition of the pressure applied on the seat/skin interface. The computer (B) enables communication between the pressure mapping system and the Tongue Display Unit (TDU). From the laptop, electrotactile stimuli can be delivered to the tongue surface via a flexible electrode array (C) placed in the mouth. This new device allows paraplegics to receive informative signals on their tongue that originate from the buttock area.

The third device is a laptop, which has been programmed to enable communication between the pressure mapping system and the TDU. From the laptop, electrotactile stimuli can be delivered to the dorsum of the tongue via a flexible electrode array (6x6), which is placed in the mouth and lightly held between the lips.

Each electrode can be activated independently. This whole device thus allows electro-stimulation of parts of the tongue according to a "signal" received from the pressure applied on the buttock area of the paraplegic. This information, that is lost after spinal cord injuries, is therefore felt again via the tongue. We hope and claim that the paraplegic will be able to develop an adapted behavior according to the electro-stimulated information in order to prevent the formation of pressure ulcers.

### 2.3. An evaluation of the feasibility of this approach.

In this preliminary work, the feasibility of the approach was evaluated by the subject's capacity to adopt an adapted behavior, with changes in pressure, according to electro-stimulated information. Our objective was therefore to demonstrate this capacity as without it, sensory substitution is of no use for this purpose. The electro-stimulated signal is not efficiently interpreted, since there is no change in pressure at the seat/skin interface, and therefore there is no possibility of relieving the cutaneous area of suffering and of preventing the formation of pressure ulcers.

After obtaining informed written consent from each subject, this evaluation was performed with 10 healthy seated subjects, at rest. For each subject, the TDU was first calibrated. This step was necessary as the perception of electrotactile stimuli on the tongue is subjective and decreases from the anterior to the posterior part of the tongue. To perform this first step, a C++ program was developed which enabled this calibration to be both user-friendly and interactive.

Following this, healthy subjects were asked to move their chest according to the felt electro-stimulated direction. More precisely, when the six electrodes placed in front of the electrode array were activated, subjects were asked to move their chest forward (Figure 2a). The instructions for movements were similar for stimulation towards the back (Figure 2b), and the left (Figure 2c) or the right sides (Figure 2d) of the electrode array.

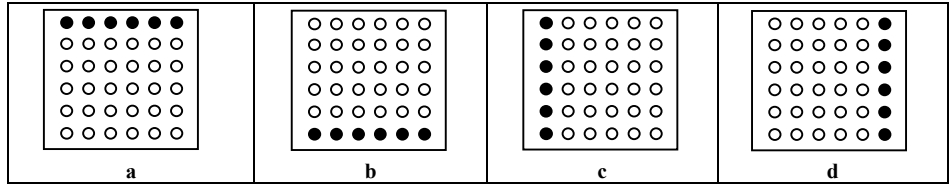

**Figure 2.** Different patterns for electro-stimulation.
When the six electrodes placed in front of the electrode array were activated, subjects were asked to move their chest forward (a). In the same way, they were asked to move their chest backward, to the left or to the right if electrodes situated behind (b), on the left (c) and on the right (d) were activated.

The pressure map applied at the seat/skin interface was then recorded in real-time. After each movement following the electro-stimulated direction, a new record of the pressure map was improved. By computing the differences between the two pressure maps (before versus after electro-stimulation), we determined whether or not the movement was adapted to the electro-stimulated information. In the first case, the result was marked as "one", and otherwise, as "zero". For each subject, this experiment was carried out 10 times, thus obtaining a total score out of 10. The entire procedure was performed in an automated manner with specifically developed software.

## 3. Results

No difficulty was reported during the calibration stage. The electro-stimulation of the tongue was very well accepted. This procedure was completed by each seated and healthy subject. The mean score was 9.2, with a standard deviation at 0.79.

This encouraging result demonstrates three main points: first, the subject has a strong perception of the electro-stimulated information; second, this information is both meaningful and correctly interpreted; and third, the action resulting from the interpreted information is adapted, with changes in pressure, to the electro-stimulatory information.

## 4. Discussion

To our knowledge, there has been no study to date that uses sensory substitution via the tongue in order to prevent the formation of pressure ulcers in paraplegics. One of the main difficulties of using sensory substitution reported by P. Bach-y-Rita was the development of a practicable user-friendly human-machine interface. Indeed, it is evident that the TDU used in this first experiment is not practicable for daily use by individuals with paraplegia. As this device has to be wholly accepted by paraplegics, it is therefore essential to solve this problem. We developed, in collaboration with the CORONIS SYSTEMS and the GUGLIELMI TECHNOLOGIES DENTAIRES companies, a new cosmetic interface incorporated into an orthodontic retainer, with wireless transmission from a laptop to the tongue device unit (Figure 3). A clinical evaluation of this prototype will start shortly. Collaboration has also been established with the VISTA MEDICAL Company in order to improve the pressure mapping system (Figure 4).

The major finding of this study is that a healthy subject is able to acquire an adapted behavior, with changes in pressure, according to electro-stimulatory information on the tongue. In other words, communication from the organ of sensory substitution towards the region of sensory loss is achieved. The reverse communication from the buttock area to the tongue has not yet been evaluated. If this reverse communication is validated, it would then enable the simulation of the whole conscious or subconscious loop defined in the healthy subject by perception of a stimulus coming from the buttock area, analysis of this signal and action adapted to the signal in order to correct the cause of this alert. The evaluation of this complete cycle will be the purpose of a further clinical study.

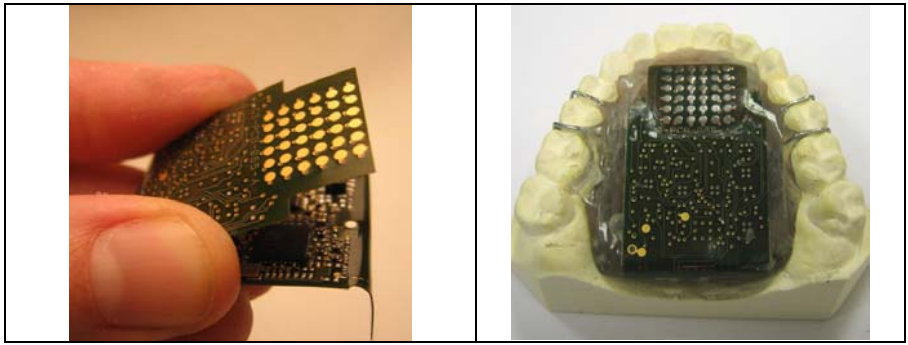

**Figure 3.** A new cosmetic interface.
On the left, the electronic interface with wireless transmission from/to the laptop is shown. On the right, this system is integrated into an orthodontic retainer, which is placed into the upper part of the mouth (bottom view).

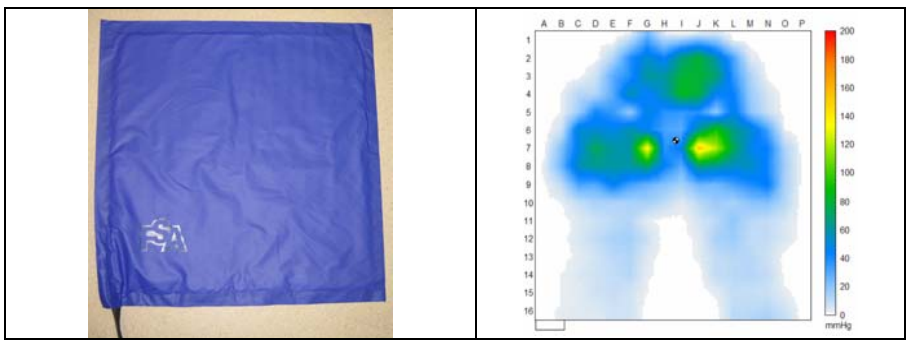

**Figure 4.** The pressure mapping system of the VISTA MEDICAL Company.
On the left, the pressure mapping system is shown. On the right, a colorized example of a pressure map at the seat/skin interface in a healthy subject is shown. In this example, the yellow and red colors correspond to the areas of maximum pressure.

Finally, it will be necessary to specify the type of information that will be applied through tongue electro-stimulation. The following questions need to be examined. Should low-level information be used as pressure that is applied to the seat/skin interface, with the subject having to interpret this information and moving in accordance with the stimuli in order to prevent the formation of pressure ulcers? Or should high-level information be used, such as an optimal direction of movement computed in an automated way from pressure maps (by minimizing a cost function of time and pressure, for example)? To obtain the best "therapeutical" acceptance, the most appropriate type of information would be the one that which would lead to a reflex-adapted behavior in the paraplegic. Further clinical evaluations should enable us to determine the most suitable information for electro-stimulation.

## 5. Conclusion

Because pressure ulcers are still a major health issue for individuals with spinal cord injuries (as shown in the high prevalence of this pathology in this population), it is necessary to develop new health strategies to address this problem. This paper reports the principles of a new approach using computer and sensory substitution via the tongue and the first performed evaluations of this technique. Initial encouraging results with healthy seated subjects, which have yet to be confirmed with paraplegics, provide the possibility of developing new strategies in health education for persons with spinal cord injuries in order to prevent the formation of pressure ulcers.